\pdfoutput=1
\documentclass[sigconf]{acmart}

\copyrightyear{2026}
\acmYear{2026}
\setcopyright{cc}
\setcctype{by}
\acmConference[RecSys '26]{20th ACM Conference on Recommender Systems}{September 27-October 02, 2026}{Minneapolis, MN, USA}
\acmBooktitle{20th ACM Conference on Recommender Systems (RecSys '26), September 27-October 02, 2026, Minneapolis, MN, USA}
\acmDOI{10.1145/3773078.3831935}
\acmISBN{979-8-4007-2284-4/2026/09}
\usepackage{booktabs}
\usepackage{multirow}
\usepackage{array}
\usepackage{xcolor}
\usepackage{tikz}
\usetikzlibrary{arrows.meta,positioning,fit,calc,backgrounds}
\usepackage{float}
\usepackage{listings}
\newcommand{\melosystem}{Melo}
\newcommand{\platform}{NetEase Cloud Music}

\title{Melo: A Production LLM-Powered Music Recommendation Agent}

\author{Shijia Wang}
\affiliation{%
  \institution{NetEase Cloud Music}
  \city{Hangzhou}
  \country{China}}
\email{wangshijia1@corp.netease.com}

\author{Da Guo}
\affiliation{%
  \institution{NetEase Cloud Music}
  \city{Hangzhou}
  \country{China}}
\email{guoda@corp.netease.com}

\author{Qiang Xiao}
\authornote{Corresponding author.}
\affiliation{%
  \institution{NetEase Cloud Music}
  \city{Hangzhou}
  \country{China}}
\email{hzxiaoqiang@corp.netease.com}

\author{Fanghui Bi}
\affiliation{%
  \institution{NetEase Cloud Music}
  \city{Hangzhou}
  \country{China}}
\email{bifanghui@corp.netease.com}

\author{Weisheng Li}
\affiliation{%
  \institution{NetEase Cloud Music}
  \city{Hangzhou}
  \country{China}}
\email{liweisheng01@corp.netease.com}

\author{Dongjing Wang}
\affiliation{%
  \institution{Hangzhou Dianzi University}
  \city{Hangzhou}
  \country{China}}
\email{Dongjing.Wang@hdu.edu.cn}

\author{Chuanjiang Luo}
\affiliation{%
  \institution{NetEase Cloud Music}
  \city{Hangzhou}
  \country{China}}
\email{luochuanjiang03@corp.netease.com}

\begin{document}

% ==== begin sections/abstract.tex ====
\begin{abstract}
We describe \melosystem{}, an LLM-powered music recommendation agent deployed on \platform{}. \melosystem{} is structured as a deterministic five-node state graph over heterogeneous tools, with a prompt- and state-machine-driven orchestration policy rather than a fine-tuned controller. At industrial scale, the bottleneck is not how smart the brain is but how the system detects and recovers from the mistakes that brain makes. Two production failure modes drove the design: entity hallucination, where the agent commits to interpretations unsupported by the live catalog or user-behavior index, and long-tail degradation, where over-constrained requests collapse to generic popular fallbacks. We address them with two complementary mechanisms. \emph{Inference-time entity grounding} repurposes the production search index as a verification primitive that gates entity decisions before they propagate downstream. \emph{Reflective retry} verbalizes failure reasons from a broken tool chain and feeds them into the next planning step, so the system can relax or revise constraints rather than fall back blindly. A one-month online A/B test across \platform{}'s playlist surfaces reports an over 2\,pp lift in a primary playlist retention metric and a lift of over one minute in a core playlist engagement metric. Offline ablation isolates a 7.8\,pp reduction in entity misidentification from the three-layer grounding stack on our evaluation set, and a triggered-session analysis on our evaluation set shows reflective retry firing on 5.8\% of sessions with 59\% process-level recovery. Our deployment experience suggests that progress on LLM-powered music recommendation at this scale depends as much on the named, ablatable runtime machinery that catches and corrects the brain's mistakes as on the brain itself: a hypothesis we offer for the community to test.
\end{abstract}
% ==== end sections/abstract.tex ====



\maketitle

% ==== begin sections/introduction.tex ====
\section{Introduction}

Production-grade natural-language music recommendation has moved from a research aspiration to a real product surface. Users now ask hundred-million-user music platforms for songs by mood, by remembered lyric fragment, by complex playlist structure, or by oblique cultural reference. They expect personalized, immediately playable catalog tracks under an interactive latency budget. Operating such a system at production scale exposes failure modes that conventional recommendation benchmarks do not see, and to our knowledge, deployment-grade experience reports of an LLM-powered music recommendation agent on a top-tier consumer surface remain limited to a single prior system~\cite{palumbo2025pfr}.

Crossing from "it works in a demo" to "it stays available for hundreds of millions of real requests" is a substantive engineering jump rather than a stronger language-understanding model. A production music agent must ground every entity it commits to in the live catalog and the user-behavior index, orchestrate heterogeneous retrieval tools under interactive latency constraints, and recover gracefully when one or more of those tools return empty or noisy results. Free-form multi-step LLM agents, while expressive, add latency variance and operational instability that are hard to justify on high-traffic product surfaces. Production music recommendation therefore needs structured orchestration with explicit failure-handling, rather than purely a stronger reasoning model.

Two high-risk failure modes recur early in deployment. The first is \emph{entity hallucination}: the agent commits to an entity whose support in the catalog or user-behavior index is implausibly weak. The bare query \emph{``Africa''} is the title of a popular Toto catalog song, yet linguistically dominated by the continent in any LLM's training distribution. An LLM-only interpretation can confidently route it to a non-music intent and surface wrong or unplayable items. The second is \emph{long-tail degradation}: complex or over-constrained requests starve one or more recall paths, and a conventional fallback-on-empty strategy discards the intermediate semantic evidence and defaults to generic popular results, silently dropping the user's core intent. Both modes are recognized as first-class problems in the agent literature: ReAct~\cite{yao2023react} on grounding, Reflexion~\cite{shinn2023reflexion} on verbalized self-correction, and recent tool-calling-agent work on infeasibility-aware constraint relaxation~\cite{awareus2025} and metric-invisible safety drift~\cite{agentdrift2025}.

The published music-recommendation literature, however, rarely treats these failures as runtime concerns with named, ablatable defenses. Spotify's Parallel Fusion Router is a single-step LLM router over predefined retrieval routes, with no explicit reflection or repair stage~\cite{palumbo2025pfr}. JD.com's hybrid multi-agent recommender focuses on search-triggering and online A/B validation in e-commerce~\cite{nie2024hybrid}. Amazon Music's playlist search and NHK's media curation use LLMs primarily for query expansion and pipeline enrichment~\cite{aluri2024playlist,hagio2025media,wang2025emotionaware}. Concurrent music-agent work either remains offline-only without deployment evidence~\cite{bi2025wemusic} or is a laboratory prototype evaluated on synthetic data~\cite{doh2025talkplaytools}. The agent literature offers reusable mechanisms such as verbal reflection and constraint-aware relaxation, but no published music recommendation system has integrated them under industrial latency budgets.

What separates a production music agent from a strong demo, then, is not the brain itself but the runtime machinery around it that catches the brain's mistakes before they reach the user. We present \melosystem{}, a deployed LLM-powered music recommendation agent serving \platform{} that treats failure detection and correction as first-class runtime mechanisms. Each defense lives at a named node in the chassis, is independently ablatable, and is evaluated against a no-defense baseline. \melosystem{} uses a deterministic five-node state graph to orchestrate heterogeneous recommendation tools through structured state transitions instead of free-form multi-step planning. This design departs from prior music agents in two ways. First, failure handling sits in named nodes that we can swap out without rewriting the controller, rather than as prompt rules in a single-step router~\cite{palumbo2025pfr} or as knowledge internalized via continued pretraining~\cite{bi2025wemusic}. Second, the same chassis hosts both existing production search and ranking services and an in-house generative retriever, letting us evaluate every runtime defense against a no-defense baseline on live traffic rather than synthetic data~\cite{doh2025talkplaytools}.

\begin{figure}[t]
  \centering
  \includegraphics[width=\columnwidth]{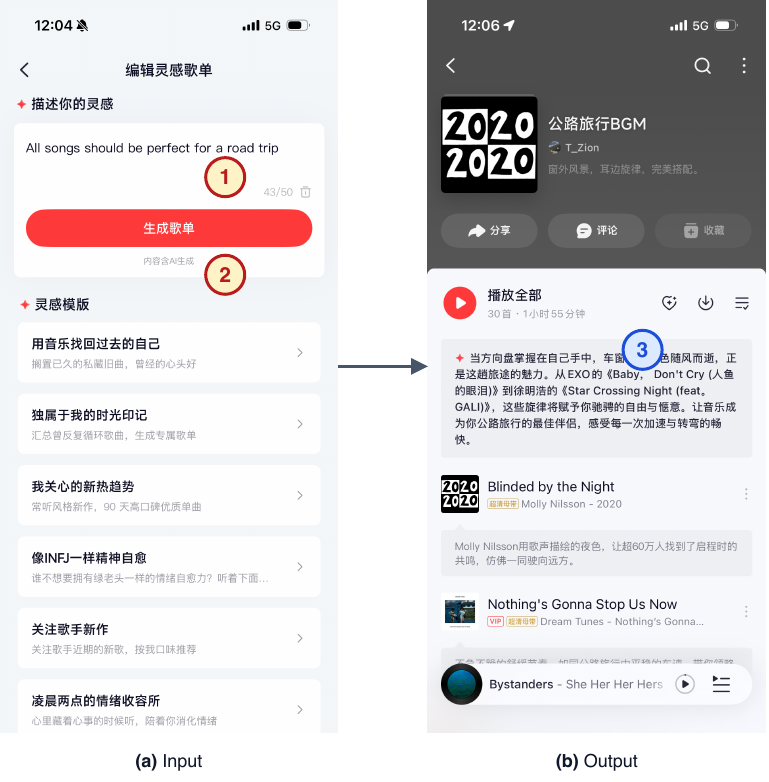}
  \caption{A real production session on \platform{}'s Muse Mix.
    The user (1)~types a natural-language music intent and
    (2)~taps generate; the system returns (3)~a complete playlist
    with AI-generated title, description, and per-track explanations
    streamed in real time.}
  \Description{Two side-by-side phone screenshots. Left: the input
    page where a user types a free-form music request. Right: the
    output page showing an AI-generated playlist with cover image,
    title, description, and song list with per-track explanations.
    Three numbered callouts mark the core flow: 1 for the intent
    input, 2 for the generate button, and 3 for the output playlist.}
  \label{fig:deployment-case}
\end{figure}

We make three contributions. First, we describe one of the few deployment-reported natural-language music recommendation agents on a hundred-million-user platform, integrated with existing recommender infrastructure as the engine behind Muse Mix (Figure~\ref{fig:deployment-case}), a natural-language playlist-generation product that has reached over one million cumulative users with strong post-launch engagement. A one-month online A/B test across \platform{}'s playlist surfaces, comparing users with Muse Mix access against users with only the existing playlist surfaces, reports measured lifts in a primary playlist retention metric and a core playlist engagement metric (Table~\ref{tab:ab}). Second, we introduce a three-layer \emph{inference-time entity-grounding} mechanism that treats the production search index as a verification primitive rather than merely a retriever. On the entity-bearing subset of our evaluation set, it reduces entity misidentification by 7.8\,pp (from 17.4\% to 9.6\%) with measured latency overhead. Third, we analyze \emph{reflective retry}, a production mechanism that verbalizes failure reasons and injects them into the next planning step. A triggered-session analysis on our evaluation set reports a 5.8\% retry trigger rate, 59\% process-level recovery, and a P50 latency overhead of 15.9 seconds.

% =====================================================================
% Figure 4: Production deployment case study (single-column, compact).
%
% Source files:
%   main/figures/drawio/figure_deployment_case_v3.drawio (editable)
%   main/figures/drawio/figure_deployment_case_v3.pdf    (rendered)
%
% Re-render after editing the drawio:
%   /Applications/draw.io.app/Contents/MacOS/draw.io \
%     -x -f pdf --crop -o figure_deployment_case_v3.pdf \
%     figure_deployment_case_v3.drawio
% =====================================================================

% ==== end sections/introduction.tex ====
% ==== begin sections/method.tex ====
\section{Methodology}

\subsection{System Design}
\label{subsec:system-design}

Our first iteration of \melosystem{} used a free-form agent template, with an LLM interleaving reasoning and tool calls in an unconstrained loop. It did not work, and the reason is informative. Production search and recommendation tools are not music's equivalent of \texttt{grep} or \texttt{bash}: they are deeply customized, business-coupled services whose argument schemas and result conventions reflect years of internal optimization rather than any uniform calling protocol. Across a free-form chain, intent-parsing, argument-extraction, and tool-side errors compounded across steps, and the resulting end-to-end success rate fell short of what we needed for production. Going back over the traces, we could segment successful requests, in hindsight, into five recurring stages: parse intent, plan tools, run them, check the result, and assemble the final response.

\begin{figure*}[t]
  \centering
  \includegraphics[width=\textwidth]{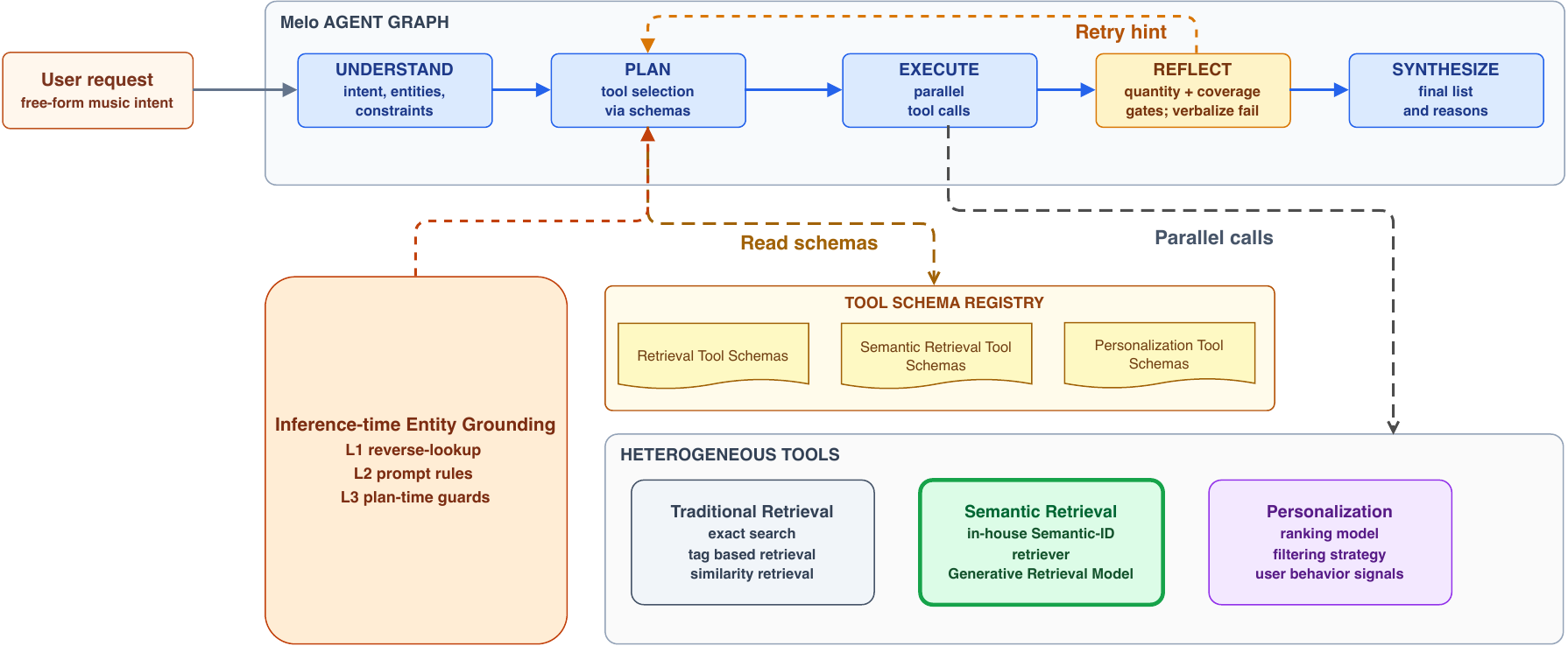}
  \caption{System overview of \melosystem{}: a deterministic
    five-node chassis (UNDERSTAND $\rightarrow$ PLAN $\rightarrow$
    EXECUTE $\rightarrow$ REFLECT $\rightarrow$ SYNTHESIZE) over a
    heterogeneous tool layer. Orange paths mark the two retrofitted
    failure-handling mechanisms---\emph{inference-time entity
    grounding} at the U/P boundary and \emph{reflective retry} on
    the REFLECT$\rightarrow$PLAN back edge.}
  \Description{A two-lane system overview. The top lane (AGENT GRAPH)
    is a five-node horizontal chassis (UNDERSTAND, PLAN, EXECUTE,
    REFLECT, SYNTHESIZE) connected by solid forward arrows; a dashed
    orange retry arrow loops from REFLECT back to PLAN labeled retry
    hint. An orange dashed inference-time entity grounding callout sits
    between the chassis and tool layer, pointing up to the
    UNDERSTAND/PLAN boundary. The bottom lane (HETEROGENEOUS TOOL
    LAYER) holds three tool boxes: Traditional Retrieval (gray),
    Climber-Think with thicker green border (in-house Semantic-ID
    retriever), and Personalization (purple). PLAN selects tools and
    EXECUTE calls them in parallel via labeled gray arrows.}
  \label{fig:system-overview}
\end{figure*}

The lesson was to make this structure explicit. \melosystem{} therefore has two layers: a deterministic five-node state graph for orchestration (Figure~\ref{fig:system-overview}), and a heterogeneous tool layer below. UNDERSTAND turns the raw request into a structured intent: entities, constraints, and personalization signals. PLAN consumes the intent and decides which tools to invoke and what arguments to pass. EXECUTE runs them in parallel and merges their candidates. REFLECT inspects the merged set and either passes it forward or sends a revised hint back to PLAN. SYNTHESIZE turns a passing set into the final track list and a short explanation. The graph topology is fixed, and LLM calls are confined to the four reasoning nodes, where they produce structured outputs and, at branch points (REFLECT$\to$PLAN vs.\ REFLECT$\to$SYNTHESIZE), drive the transition choice. Compared with ReAct-style loops~\cite{yao2023react}, this trades flexibility for two properties a hundred-million-user surface needs: failures can be attributed to a specific node, and defenses can be added or replaced at a single node without changing the rest of the graph. We refer to these five stages, each with a stable name and a typed input/output contract, as \emph{named nodes}, and a defense attached to exactly one of them as a \emph{named-node defense}.

% =====================================================================
% Figure 1: Melo system overview.
%
% Source files:
%   main/figures/drawio/figure1_system_overview.drawio (editable)
%   main/figures/drawio/figure1_system_overview.pdf    (rendered)
%
% Re-render after editing the drawio:
%   /Applications/draw.io.app/Contents/MacOS/draw.io \
%     -x -f pdf --crop -o figure1_system_overview.pdf \
%     figure1_system_overview.drawio
% =====================================================================

The tool layer mixes two kinds of components: existing production services such as catalog search, entity grounding, profile and behavior retrieval, candidate generation, similarity retrieval, and personalized ranking, alongside an in-house generative retriever, which maps free-form requests to playable catalog tracks via Semantic IDs and builds on the team's prior work on residual and dense quantization~\cite{wang2025psrq,xiao2026denserq,xu2025climber,guo2026climberpilot}. Owning both layers lets us co-tune them on live traffic, and the orchestration layer's job is to decide when each retrieval surface fits the current intent.

Even with these two layers in place, the chassis is not a guarantee: LLM-driven transitions and tool inconsistencies still compound into two production failures the chassis alone cannot eliminate. An early LLM decision can lock in an entity interpretation the live catalog does not support: this is the entity-hallucination mode from the introduction. A retrieval pass can return empty or constraint-incompatible candidates that a naive fallback collapses to generic popular results: this is the long-tail-degradation mode. We attach two failure-handling mechanisms at specific nodes of the chassis: inference-time entity grounding on UNDERSTAND/PLAN (\S\ref{subsec:entity-grounding}) and reflective retry from REFLECT back to PLAN (\S\ref{subsec:reflective-retry}), the latter capped at two rounds by the latency budget. These two failure modes are the most consequential ones we have documented. Standard infrastructure failures such as network errors, tool API changes, and timeouts are handled by orthogonal retry and timeout policies outside this paper's scope. Both mechanisms were retrofitted once production exposed the failures, and that each could be added at a named node without rewriting the graph is itself an argument for the chassis design. Underlying that argument is a deliberate bet about what matters at this scale: the runtime scaffolding around the brain, which surfaces and corrects its mistakes at named nodes rather than in prompts, is at least as discriminating as the brain itself.

\subsection{Inference-Time Entity Grounding}
\label{subsec:entity-grounding}

The first failure-handling mechanism attached to the chassis sits between UNDERSTAND and PLAN. Recall the ``Africa'' example from the introduction. UNDERSTAND must decide whether the user means the Toto song, an artist named Africa, or (most likely in any LLM's training distribution) the continent. The mistake we kept seeing in production was not a retrieval mistake but an interpretation mistake that retrieval then dressed up as a confident answer.

\begin{figure}[t]
  \centering
  \begin{tikzpicture}[
    node distance=5mm,
    box/.style={draw, rounded corners, align=center, minimum width=38mm, minimum height=8mm, fill=orange!6},
    arr/.style={-Latex, thick}
  ]
    \node[box] (l1) {L1 Catalog reverse-lookup\\support signals};
    \node[box, below=of l1] (l2) {L2 Prompt consumption rules\\weigh evidence};
    \node[box, below=of l2] (l3) {L3 Plan-time guards\\revise or verify};
    \node[box, below=of l3, fill=green!6] (out) {Grounded plan};
    \draw[arr] (l1) -- (l2);
    \draw[arr] (l2) -- (l3);
    \draw[arr] (l3) -- (out);
  \end{tikzpicture}
  \caption{Inference-time entity grounding: three layers progressively verify whether a candidate entity interpretation is supported by the live catalog before the plan commits to it.}
  \Description{A three-layer verification stack: catalog reverse-lookup provides support signals, prompt consumption rules weigh the evidence, and plan-time guards revise or verify before producing a grounded plan.}
  \label{fig:search-verification}
\end{figure}
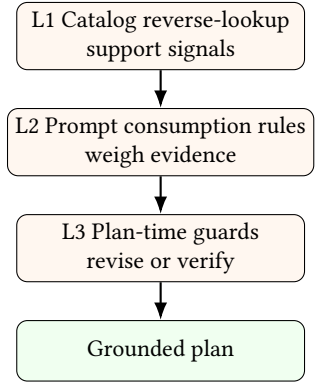

Our inference-time entity-grounding stack turns this from a prompt problem into a runtime check, treating the production search index as a verification primitive rather than as a retriever. Three layers of entity-grounding act on the candidate interpretation as it flows from UNDERSTAND into PLAN, and each answers a different question. The first layer (L1: catalog reverse-lookup) asks whether the catalog actually supports this interpretation: a production reverse-lookup service takes each candidate entity, queries the catalog and search index, and returns support signals such as entity type, catalog availability, and coarse popularity or activity indicators, so the agent sees evidence rather than just a name. The second layer (L2: prompt-level consumption rules) asks how the agent should weigh that evidence: prompt-level consumption rules in UNDERSTAND tell the LLM how to consume the reverse-lookup output, deciding when low activity should downweight a confident-looking name, when an ambiguous short token should be flagged rather than silently resolved, and when a popularity signal is too thin to commit to a planned tool call. The third layer (L3: plan-time guards) asks what the planner should do if support is still weak: plan-time guards in PLAN intercept structured intents that match recurring failure patterns such as explicit negations, verb-prefix false matches, or unverified artist identifiers, and force a revision, a broader query, or a safer retriever before any tool that would amplify the wrong answer is allowed to fire.

Entity grounding in our sense, catching a misinterpretation before it propagates into tool calls, is a different problem from \emph{catalog grounding} as used in generation-time recommenders, which maps an LLM's already-emitted item name back onto the catalog~\cite{zhu2026rankgrpo}. It is also more than classical entity linking: where entity linking produces a best-mapping decision in isolation, our stack additionally gates downstream tool calls on link confidence, so that a misinterpretation that linking would tolerate as ``best available'' is rejected here as ``too weak to act on''. Section~\ref{sec:anti-hallucination} reports a layer-wise ablation across the three layers and a no-defense baseline.

\subsection{Reflective Retry}
\label{subsec:reflective-retry}

The second failure-handling mechanism closes the loop from REFLECT back to PLAN. The long-tail-degradation mode happens when a request makes it through UNDERSTAND and PLAN cleanly but EXECUTE returns too few candidates, usually because a recall path missed an in-catalog entity, a tool hit a coverage limit, or the request stretched across a sparsely-cataloged region. Our first response was the standard popular-list backup, but in production this dropped the user's intent often enough to be a real problem.

Reflective retry replaces the silent fallback with a verbalized one. REFLECT applies two gates to EXECUTE's merged set: a minimum-size check and a coverage check against the original intent. When the merged set fails either gate, REFLECT, itself an LLM call, selects one of four actions: \texttt{proceed}, when the merged set is sufficient and the agent moves to SYNTHESIZE; \texttt{retry\_rewrite}, when the structured intent itself was off and PLAN should revise it, for instance by correcting a wrong entity binding or adding a missing slot; \texttt{retry\_relaxed}, when the intent is correct but a constraint must be loosened to broaden coverage; or \texttt{fallback}, when the retry budget is exhausted and the system surfaces a partial result with an explanation. On a retry action, REFLECT verbalizes the suspected cause and one suggested revision and feeds them back to PLAN as additional input on the next planning step. PLAN then issues a revised tool chain: relaxing one or more constraints, splitting a compound intent into separate retrieval routes, or switching to a recall path more tolerant of long-tail items. This is the verbal-feedback idea of Reflexion~\cite{shinn2023reflexion} compressed into a single, latency-bounded session, and the loop runs at most twice.

Figure~\ref{fig:retry-trace} shows a real production trace on the query \textit{``niche Icelandic post-punk''}: Round 1's single tag query with both region (Iceland) and genre (post-punk) constraints returned only 3 candidates; REFLECT diagnosed both as simultaneously over-tight and asked PLAN to relax region from Iceland to Nordic, expand genre to post-punk plus alternative rock, and add a semantic-recall path targeting `cold Icelandic atmosphere'; Round 2 returned 42 candidates and cleared both gates.

\begin{figure}[t]
\small
\hrule height 0.5pt
\vspace{4pt}
\begin{flushleft}
\ttfamily
PRODUCTION QUERY:~~``niche Icelandic post-punk''
\end{flushleft}
\vspace{2pt}
\hrule height 0.4pt
\begin{flushleft}
\ttfamily
\textbf{Round 1 plan} \\
1 retrieval dispatched: \\
\hspace*{1em}[1] tag search\hspace*{2.4em}region=Iceland, genre=post-punk \\[2pt]
EXECUTE returned 3 candidates.
\end{flushleft}
\vspace{2pt}
\hrule height 0.4pt
\begin{flushleft}
\ttfamily
\textbf{REFLECT (Round 1)} \\
\hspace*{1em}candidates returned : 3 \\
\hspace*{1em}sufficient\hspace*{10pt}~~: no \\
\hspace*{1em}diagnosis\hspace*{12pt}~: ``Geography and genre constraints \\
\hspace*{1em}\hspace*{8em}~~are simultaneously over-tight.'' \\
\hspace*{1em}hint\hspace*{34pt}~~: ``Relax geography to Nordic; expand \\
\hspace*{1em}\hspace*{8em}~~genre to post-punk + alternative \\
\hspace*{1em}\hspace*{8em}~~rock; add a semantic-recall path \\
\hspace*{1em}\hspace*{8em}~~covering `cold Icelandic atmosphere'.'' \\
\hspace*{1em}next action\hspace*{4pt}~~: retry\_relaxed
\end{flushleft}
\vspace{2pt}
\hrule height 0.4pt
\begin{flushleft}
\ttfamily
\textbf{Round 2 plan} (revised per hint) \\
3 parallel retrievals dispatched: \\
\hspace*{1em}[1] tag search\hspace*{2.4em}region=Nordic, genre=post-punk \\
\hspace*{1em}[2] tag search\hspace*{2.4em}region=Nordic, genre=alt-rock \\
\hspace*{1em}[3] semantic recall\hspace*{0.4em}``cold Icelandic atmosphere'' \\[2pt]
REFLECT $\to$ proceed.~~Returned a 15-track playlist.
\end{flushleft}
\vspace{2pt}
\hrule height 0.5pt
\caption{A reflective-retry trace from production.}
\label{fig:retry-trace}
\Description{A four-block trace of a single production session for the query "niche Icelandic post-punk": Round 1 plan with one tag-based retrieval combining region and genre filters returning only 3 candidates; the REFLECT block showing candidate count, sufficiency, a diagnosis identifying both constraints as simultaneously over-tight, a hint to broaden geography to Nordic, expand genre to post-punk plus alternative rock, and add a semantic-recall path covering 'cold Icelandic atmosphere', and next action retry_relaxed; Round 2 plan with three parallel retrievals revised per the hint returning 42 candidates; and a final line indicating REFLECT proceed and a playlist with 42 tracks.}
\end{figure}

Retry must not hide what it gives up. In early diagnostic runs, a heavily over-constrained query recovered from zero candidates to a non-empty result, but some of the original constraints had quietly been dropped on the way. This is a real failure shape, not an edge case, and it tells us that production retry has to be evaluated by two metrics in tension: candidate recovery and constraint preservation. Section~\ref{sec:long-tail} reports both, with constraint preservation called out as an open evaluation question.

\subsection{Implementation Choices}
\label{subsec:impl-choices}

Two implementation choices round out the deployment. The first concerns ownership: \melosystem{} keeps trained recommendation models behind stable tool interfaces and drives orchestration with prompts and a state machine, so model teams can update retrieval and ranking while the agent team hot-updates prompts, tool schemas, and transitions---neither side retraining a policy model.

The second concerns latency: two techniques keep end-to-end response within the interactive budget. First, tool descriptions are loaded on demand, so PLAN's prompt only carries the skills relevant to the current intent rather than the full toolbox. Second, SYNTHESIZE streams intermediate progress and final results through server-sent events, so the user starts seeing the response while ranking and explanation generation are still running.
% ==== end sections/method.tex ====
% ==== begin sections/experiments.tex ====
\section{Experiments}

\subsection{Setup}

We evaluate \melosystem{} along three axes: production impact, failure-specific effectiveness, and operational cost. Production impact is measured by a one-month online A/B test across \platform{}'s playlist surfaces: the treatment arm sees a Muse Mix entry alongside the existing playlist surfaces, and the control arm sees only the existing surfaces. Failure-specific effectiveness is measured by two targeted analyses on an evaluation set (described next): a layer-wise ablation of the entity-grounding stack and a triggered-session analysis of reflective retry. Each analysis is run three times with Kimi K2.5~\cite{kimiteam2026kimik25} as the backbone LLM at temperature 0.3 to surface LLM non-determinism, and we report mean $\pm$ one standard deviation. Additional evaluations with DeepSeek-V4-Flash yielded consistent directional conclusions; the production system runs a multi-model configuration for operational stability. Operational cost is measured by node-level latency and incremental overhead introduced by verification and retry.

\textbf{Evaluation set.} We construct an internal evaluation set of 298 queries by stratified sampling from a large corpus of real typing queries on \platform{}, organized by query type and supplemented with a few templated probes for failure patterns whose natural production frequency is too low to yield a stable signal. Because it deliberately over-samples the two failure modes this paper targets (entity ambiguity and long-tail/complex requests), the set is a targeted stress test rather than a traffic-proportional sample; rates measured on it (e.g., the 5.8\% retry-trigger rate) are properties of this set, not estimates of raw production frequency. The set is stratified into three classes, each tied to a specific analysis below:
\begin{itemize}
  \item \emph{Entity-ambiguity queries} (\textasciitilde35\%): title--artist homonyms, ``A's B'' constructions, verb-prefix commands such as ``play X'', negation patterns, alias or nickname references, and artist queries phrased like tag queries. Primary target of the entity-grounding ablation.
  \item \emph{Long-tail or complex queries} (\textasciitilde40\%): multi-constraint cross, complex negation, niche artists, cross-era or cross-genre composites, abstract emotion or scene, lyric-content retrieval, and mixed precise-plus-fuzzy intents. Primary driver of retry triggers in the reflective-retry analysis.
  \item \emph{Regular baseline queries} (\textasciitilde25\%): exact title or artist lookup, tag and emotion recommendation, and history-based personalization. The no-failure-mode floor for both analyses and the dominant pool of non-retry sessions used in the latency report.
\end{itemize}
Every query carries a design intent and reference annotations along four dimensions---entity correctness, retry behavior, failure mode, and result quality---produced under a unified annotation guideline. Each query is replayed three times under each configuration with backbone LLM temperature 0.3 to surface stochasticity, and the per-configuration measurement volume is on the order of $10^3$. The set is small relative to the platform's live traffic and covers Chinese music-platform queries only; larger, cross-market evaluation remains future work.

\subsection{Online A/B Test}

\begin{table}[t]
  \caption{Online A/B test on \platform{}'s playlist surfaces; values are conservative lower bounds (\texttt{>X}).}
  \label{tab:ab}
  \centering
  \begin{tabular}{lc}
    \toprule
    Metric & Reported impact \\
    \midrule
    Primary playlist retention metric      & \textbf{$>$2\,pp} \\
    Core playlist engagement metric        & \textbf{$>$1\,min} \\
    \bottomrule
  \end{tabular}
\end{table}

Table~\ref{tab:ab} reports the online A/B result. We compared two randomized arms across \platform{}'s playlist surfaces from April 2 to May 10, covering approximately one million users with equal traffic allocation between arms: a treatment arm in which users have access to Muse Mix as an additional playlist surface alongside the existing playlist products, and a control arm in which users have only the existing surfaces. Two surface-level metrics moved in the treatment arm. A primary playlist retention metric which is defined in Table~\ref{tab:ab} rose by over 2 percentage points. A core playlist engagement metric, also aggregated across all playlist surfaces, rose by over one minute. We read the two surface-level lifts together as consistent with expansion rather than cannibalization: a Muse Mix-internal lift would have been the natural prediction, while a lift across all playlist surfaces means the new surface attracts behavior that the existing surfaces were not capturing. Because Muse Mix is a product surface, this experiment measures the value of the deployed system as a whole: \melosystem{} exists to power Muse Mix and cannot be stripped of its product shell for an isolated online test. The two mechanism-level contributions are therefore evaluated separately and offline---the layer-wise entity-grounding ablation (\S\ref{sec:anti-hallucination}) and the paired retry-on/off ablation (\S\ref{sec:long-tail})---rather than attributed to this A/B. We do not report confidence intervals or significance tests, and one-month novelty effects, within-arm substitution, and the product-surface/UI change are not fully ruled out by this experiment.

\subsection{Anti-Hallucination Ablation}
\label{sec:anti-hallucination}

\begin{table}[t]
\centering
\caption{Anti-hallucination ablation: removing defense layers progressively increases entity misidentification while reducing latency.}
\label{tab:ablation}
\small
\setlength{\tabcolsep}{2pt}
\begin{tabular}{lcccc}
\toprule
Configuration & \shortstack{Misidentification \\ rate $\downarrow$} & $\Delta$ & \shortstack{UNDERSTAND \\ P50} & \shortstack{PLAN \\ P50} \\
\midrule
Full Defense (L1+L2+L3)   & 9.6\%\,$\pm$\,2.1\%  & ---      & 9.6\,s & 9.7\,s \\
$-$L3 (no guards)         & 10.8\%\,$\pm$\,0.0\% & +1.2\,pp & 8.9\,s & 9.2\,s \\
$-$L2$-$L3 (L1 only)      & 13.4\%\,$\pm$\,0.8\% & +3.8\,pp & 8.3\,s & 8.9\,s \\
No Defense                & 17.4\%\,$\pm$\,1.8\% & +7.8\,pp & 7.0\,s & 7.9\,s \\
\bottomrule
\end{tabular}
\end{table}

Table~\ref{tab:ablation} reports the ablation. We replay four configurations (full defense, $-$L3, $-$L2$-$L3, and no defense) against the entity-bearing subset of the evaluation set, three times each. We exclude queries with intermittent catalog reverse-lookup unavailability across all three full-defense runs and queries flagged as genuinely ambiguous. If unavailability correlates with entity complexity, the remaining subset skews toward easier queries, so the reported delta is a conservative lower bound on the population effect.

Removing each layer measurably increases misidentification, with unequal contributions. Removing plan-time guards alone (L3) raises misidentification by 1.2\,pp, which is small but consistent across runs because the guards mainly catch recurring tail patterns such as explicit negation and verb-prefix false matches rather than the head distribution. Removing the prompt consumption rules on top (L2) adds another 2.6\,pp, the largest single-layer contribution: catalog evidence alone is not enough, since the LLM also needs explicit instruction on how to weigh popularity, activity, and entity-type signals before committing. Removing the catalog reverse-lookup as well (L1) yields a baseline of 17.4\,\% misidentification, 4.0\,pp above the catalog-only configuration.

The defense layers cost latency. Full defense adds about 2.5\,s to the UNDERSTAND P50 and 1.8\,s to the PLAN P50 over the no-defense baseline, roughly the cost of one additional reverse-lookup round trip plus a longer prompt. Since most wrong cases at the no-defense baseline are empty outputs rather than confidently wrong entities, the defense layers protect against both silent failure (entity resolver returning nothing) and confident misidentification, in roughly that order. Together with the $\sim$8\,pp misidentification reduction, this is the empirical backing for our second contribution: inference-time entity grounding cuts misidentification by 7.8\,pp as a runtime, ablatable mechanism, not a prompt-engineering artefact.

\subsection{Long-Tail Coverage and Reflective Retry}
\label{sec:long-tail}

\begin{table}[t]
\centering
\caption{Triggered-session view of reflective retry.}
\label{tab:retry}
\small
\begin{tabular}{lr}
\toprule
\textbf{Metric} & \textbf{Value} \\
\midrule
Retry trigger rate                   & 5.8\% \\
\quad via \texttt{retry\_relaxed}         & 52\% \\
\quad via \texttt{retry\_rewrite}         & 48\% \\
Retry success rate (overall)         & 59.1\% \\
\quad \texttt{retry\_relaxed}             & 63\% \\
\quad \texttt{retry\_rewrite}             & 56\% \\
Constraint-preservation (recovered)  & 77\%\\
Extra latency (P50)                  & +15.9\,s \\
Extra latency (P99)                  & +29.2\,s \\
  \bottomrule
\end{tabular}
\end{table}

Table~\ref{tab:retry} reports a triggered-session view: across three replays of the evaluation set under the production configuration, the REFLECT-to-PLAN back edge fires on roughly 5.8\% of sessions, and 59.1\% of those recover by REFLECT eventually returning \texttt{proceed} while the remaining 40.9\% exhaust the two-round budget. Counted by the action that initiates the retry, the trigger splits almost evenly between \texttt{retry\_relaxed} (52\% of triggered sessions, recovering 63\% of the time) and \texttt{retry\_rewrite} (48\%, recovering 56\%). This split matters for reading the success rate: \texttt{retry\_rewrite} repairs a \emph{misread} intent, while \texttt{retry\_relaxed} deliberately \emph{loosens} an over-tight constraint, so part of what the code counts as recovery is intended relaxation rather than a constraint-faithful match. Not every failure is recoverable: when the agent has already misunderstood the request or the requested entity is genuinely absent from the catalog, rewriting or relaxing addresses the wrong problem.

Retry costs latency. Triggered sessions take 15.9\,s longer at P50 and 29.2\,s longer at P99 than non-retry sessions, roughly the cost of an extra LLM call plus a second tool round. Because retry fires on only 5.8\% of sessions, the amortized cost over all traffic is just +4.1\,s at P50 (Table~\ref{tab:retry_onoff}). These are sessions that would otherwise have returned empty results, so the relevant comparison is not ``fast vs.\ slow'' but ``slow but useful vs.\ fast and empty.''

\begin{table}[t]
\centering
\caption{Retry-on vs.\ retry-off paired ablation on the evaluation set under matched configuration.}
\label{tab:retry_onoff}
\small
\setlength{\tabcolsep}{4pt}
\begin{tabular}{lrrr}
\toprule
Metric & Retry-on & Retry-off & $\Delta$ \\
\midrule
Avg.\ song count     & 18.45 & 18.24 & +0.21 \\
P50 latency                   & 32.5\,s & 28.4\,s & +4.1\,s \\
P99 latency                   & 57.2\,s & 43.6\,s & +13.6\,s \\
\bottomrule
\end{tabular}
\end{table}

To isolate what retry's absence would cost, we ran a paired retry-on / retry-off ablation across the full evaluation set (Table~\ref{tab:retry_onoff}). On the queries it materially improves, retry-on adds an average of +17.5 candidate songs over the retry-off baseline. The average song-count delta over all queries is +0.21. The full-traffic amortized latency adds +13.6\,s at P99 (the +4.1\,s at P50 noted above). The asymmetry is the point: retry is cheap on the median session because it does not run, and high-leverage on the tail sessions where it does.

In our manual review of production failure traces, retry reliably recovers sessions where the first-round retrieval came back too sparse to satisfy the request: over-constrained queries that a relaxed plan can reach, or narrow regions that a broader recall path can cover. Two cases illustrate the dual outcomes. The query \textit{``niche Icelandic post-punk''} is a cleanly recovered case (full trace in Figure~\ref{fig:retry-trace}). A query for \textit{``1960s Mongolian-language rock by female singers, slow tempo''} is a recovered-but-degraded case: the five stacked constraints return zero candidates, so REFLECT issues \texttt{retry\_relaxed}, and the relaxed plan recovers a full twenty-track playlist that keeps the dominant intent of \emph{Mongolian-flavored rock}---for instance songs by Tengger (a well-known Mongolian rock vocalist) and Tang Dynasty (a pioneering rock band), alongside morin-khuur (horse-head-fiddle) instrumentals---while silently dropping the narrower 1960s, Mongolian-\emph{language}, female-vocalist, and slow-tempo constraints. Crucially, semantic \emph{degradation} of this kind is not the same as intent \emph{violation}: on a production recommendation surface, a degraded but on-theme, immediately playable result is more useful than a constraint-faithful empty one. To quantify this, we manually audited the recovered sessions: 77\% preserve the core constraint and 23\% are recovered-but-degraded, with degradation concentrated in \texttt{retry\_relaxed} 29.4\% rather than \texttt{retry\_rewrite} 14.3\%; intent-preserving recovery is therefore roughly 46\% of triggered sessions. We report this constraint-preservation rate separately rather than folding it into the process-level 59.1\%. This asymmetry is what makes reflective retry our third contribution rather than a generic try-again loop: the same mechanism that costs 15.9\,s on the 5.8\% triggered tail converts otherwise-empty sessions into usable ones, while leaving the 94.2\% median path untouched.

\subsection{Latency and Cost}
\label{sec:latency}

\begin{table}[t]
  \caption{Node-level P50/P99 latency on non-retry sessions, full-defense configuration; three replays of the evaluation set.}
  \label{tab:latency}
  \centering
  \begin{tabular}{lcc}
    \toprule
    Stage & P50 latency & P99 latency \\
    \midrule
    UNDERSTAND        & 8.0\,s  & 18.6\,s \\
    PLAN              & 8.2\,s  & 18.8\,s \\
    EXECUTE           & 0.04\,s & 0.5\,s  \\
    REFLECT$^\dagger$ & 0.0\,s  & 3.9\,s  \\
    SYNTHESIZE        & 10.3\,s & 23.4\,s \\
    End-to-end        & 26.0\,s & 52.6\,s \\
    \bottomrule
  \end{tabular}
  \vspace{2pt}
  \begin{flushleft}
  \footnotesize
  $^\dagger$REFLECT P50 is near zero because for non-retry sessions the node fast-paths once the first-round merge passes the coverage gate; on the 5.8\,\% of triggered sessions REFLECT issues an additional LLM call with P50 6.4\,s.
  \end{flushleft}
\end{table}

Latency is a first-order constraint, but the question for an LLM-powered agent is where the budget goes. Table~\ref{tab:latency} reports node-level P50 and P99 for non-retry sessions under full defense. LLM reasoning, not tool dispatch, dominates: UNDERSTAND, PLAN, and SYNTHESIZE collectively account for nearly the entire 26.0\,s end-to-end median, while EXECUTE and REFLECT sit below 100\,ms at the median. This pattern both validates the parallel-execute design and identifies the next optimization frontier. P99 follows the same shape: SYNTHESIZE and PLAN widen the tail, whereas EXECUTE stays under 0.6\,s because tool-side variance is masked by parallelism. REFLECT's median is an artefact of the non-retry filter. On the 5.8\,\% of triggered sessions REFLECT itself takes 6.4\,s at P50 because of the additional LLM call, with the round-trip cost reported in \S\ref{sec:long-tail}. Muse Mix is an asynchronous playlist-generation surface with server-sent-event streaming, so the user begins seeing tracks well before the full-generation median of 26.0\,s; the node-level breakdown in Table~\ref{tab:latency} localizes where that budget is spent. We treat sub-second \emph{synchronous} conversational use as a different product target and out of scope here.
% ==== end sections/experiments.tex ====
% ==== begin sections/discussion.tex ====
\section{Discussion}

\textbf{Cost: keep prompts tight or pay at every node.} \melosystem{} dispatches an LLM call at every reasoning node, and naive prompts quickly inflated per-call token counts to the point of undermining the product's unit economics. Three practices brought cost back into a sustainable range: node-specific prompts that carry forward only what the next node needs, on-demand tool descriptions in PLAN (rather than the full toolbox in every prompt), and prompt caching for slow-varying segments such as system instructions, tool schemas, and catalog snapshots. None of these is novel in isolation. What mattered for production was treating them as a discipline rather than a nice-to-have.

\textbf{LLM stochasticity is a system property, not a quirk.} The same prompt and the same input do not produce the same output across calls, and that variance compounds along a chain of LLM-driven decisions. Our concrete lesson is that any logic that can be made deterministic should be. Code-level plan-time guards (L3), schema-validated structured outputs at every reasoning node, and the explicit four-action REFLECT enum all exist because we found that delegating the same decision to an LLM repeatedly was both slower and less stable than codifying it once. The art is knowing where to keep flexibility, such as natural-language understanding and free-form reasoning over partial evidence, and where to lock things down.

\textbf{Latency is a moving target.} Single-session generation time is uncomfortably long for an interactive surface, but this is not specific to \melosystem{}: every LLM-powered agent we are aware of carries comparable overhead, and we treat it as a near-term problem rather than a structural one. Two trends will pull it down. Because UNDERSTAND, PLAN, and SYNTHESIZE dominate the budget while EXECUTE/REFLECT are sub-100\,ms, the optimization vectors are node-anchored: (i) distilled or flash-tier models for the three reasoning nodes (each model generation has already cut our P50 by a measurable margin without code changes); (ii) prompt compression and caching of slow-varying segments; (iii) non-LLM short-circuits for high-frequency head intents, reserving full LLM reasoning for the genuinely novel tail; and (iv) further parallelizing and streaming SYNTHESIZE.

\textbf{What does and does not generalize.} The chassis design and the failure-handling pattern are general, but several pieces of \melosystem{} are platform-specific. The entity-grounding thresholds, such as popularity cutoffs and ambiguity heuristics, reflect distributions on \platform{} and would need to be re-tuned on a different catalog. The retry budget and its latency arithmetic are tied to an SSE-streaming surface where the user is already seeing partial output, so an agent without streaming would face a different cost--benefit calculation. We did not isolate reflective retry in a dedicated A/B test because its 5.8\% trigger rate is below the resolution of aggregate product metrics, so we evaluate it through triggered-session analysis instead. Cross-market and cross-catalog validation remains future work. Finally, \melosystem{}'s retrieval claim is coverage recovery---avoiding empty results and the collapse to generic popular fallbacks---not diversity or novelty optimization; explicit item- and user-level diversity/novelty metrics, and a comparison against dedicated diversity-optimizing or matrix-factorization rankers, are out of scope for this coverage-focused report and left to future work.
% ==== end sections/discussion.tex ====
% ==== begin sections/related_work.tex ====
\section{Related Work}

\textbf{LLM agents for recommendation.}
The ``LLM as brain, recommender as tools'' paradigm has developed rapidly since 2023, as surveyed in~\cite{zhang2025surveyagents}.
RecMind~\cite{wang2024recmind} first demonstrated zero-shot recommendation via LLM planning over external database and search tools. AgentCF~\cite{zhang2024agentcf} introduced bilateral user--item agent modeling with collaborative reflection.
InteRecAgent~\cite{huang2025interecagent} proposed plan-then-execute workflows across non-music domains. DualAgent-Rec~\cite{zhang2026dualagent} deploys LLMs as orchestrators over constraint-aware evolutionary optimization for e-commerce.
Recent work trains tool-calling policies: RecThinker~\cite{zhang2026recthinker} uses self-augmented SFT and GRPO for Analyze-Plan-Act reasoning, and STAR~\cite{wu2026star} distills multi-agent traces into a single student model.
Rank-GRPO~\cite{zhu2026rankgrpo} trains conversational recommenders via RL with catalog-grounding verification, a generation-time mechanism complementary to our inference-time entity-grounding stack.
These systems are evaluated on academic benchmarks and do not address production constraints such as latency budgets or tool service unreliability.

\textbf{Music and production systems.}
Parallel Fusion Router~\cite{palumbo2025pfr} from Spotify, the closest deployed system to ours, is a single-step LLM router post-trained for latency and validated via large-scale online A/B; its latency-first design forgoes the reflection-and-retry stage that \melosystem{} pays for failure-recoverability. Other music systems stop short of this setting: TalkPlay-Tools~\cite{doh2025talkplaytools} is a synthetic-data prototype, WeMusic-Agent~\cite{bi2025wemusic} internalizes knowledge via continued pretraining without reported online A/B, and Amazon Music~\cite{aluri2024playlist} and NHK~\cite{hagio2025media} use LLMs for query expansion without agentic loops. In adjacent domains, JD.com's hybrid CRS~\cite{nie2024hybrid}, SIGMA~\cite{yu2026sigma} at AliExpress, and Netflix's log-verbalization work~\cite{shi2025fromlogstolanguage} report production generative/agentic recommendation with online A/B.
% ==== end sections/related_work.tex ====
% ==== begin sections/conclusion.tex ====
\section{Conclusion}

We presented \melosystem{}, a production LLM-powered music recommendation agent built around a five-node state graph. The system treats failure detection and correction as first-class production requirements through two ablatable defense mechanisms attached to named chassis nodes: inference-time entity grounding and reflective retry. A one-month online A/B test across \platform{}'s playlist surfaces shows that adding Muse Mix lifts a primary playlist retention metric by over 2\,pp and a core playlist engagement metric by over one minute, aggregated across all playlist surfaces. Offline ablation isolates a 7.8\,pp reduction in entity misidentification from the three-layer grounding. A triggered-session analysis on our evaluation set shows reflective retry achieving 59\% process-level recovery on the 5.8\% of sessions where it fires. More broadly, our deployment experience suggests that at this scale, an LLM-powered music agent's headroom comes less from a stronger brain than from the named, ablatable apparatus around it that intercepts the brain's mistakes and turns them into recoverable signals.
% ==== end sections/conclusion.tex ====

\begin{acks}
This research was supported by the Natural Science Foundation of Zhejiang Province under Grant No.~LZ25F020010.
\end{acks}

\bibliographystyle{ACM-Reference-Format}
\bibliography{references}

\appendix
% Appendix-scoped tightening to respect the 9-page limit.
\lstset{basicstyle=\ttfamily\scriptsize,aboveskip=2pt,belowskip=1pt}
\setlength{\textfloatsep}{4pt plus 1pt minus 1pt}
\setlength{\intextsep}{4pt plus 1pt minus 1pt}
\setlength{\abovecaptionskip}{2pt}
\setlength{\belowcaptionskip}{2pt}
\section{Reproducibility Details}
\label{app:repro}

\textbf{REFLECT node contract.}
REFLECT emits a typed object with the fields and action enum shown below. A fast rule gate short-circuits to \texttt{proceed} (without an LLM call) when both hold: candidate count $\ge \tau_{\text{size}}\!\cdot\!N$ ($\tau_{\text{size}}{=}1.5$, target size $N{=}20$) and at least one recall route returned at least one result; otherwise an LLM decides the action. Personal-set requests (liked/listened) are hard-protected to \texttt{proceed}. Retry is capped at two rounds; the verbalized failure reason and \texttt{relaxed\_params} are fed back to PLAN.
\begin{lstlisting}
{
  "is_sufficient": "bool",
  "total_candidates": "int",
  "semantic_match_score": "float in [0,1]",
  "action": "proceed | retry_rewrite | retry_relaxed | fallback",
  "relaxed_params": "object (fed back to PLAN)",
  "corrected_understanding": "object | null",
  "reasoning": "string"
}
\end{lstlisting}

\textbf{Three-layer entity grounding.}
Table~\ref{tab:grounding} summarizes the three ablatable defense layers.

\begin{table}[H]
\caption{Three-layer inference-time entity grounding (each layer is independently ablatable).}
\label{tab:grounding}
\centering\scriptsize\renewcommand{\arraystretch}{0.95}
\begin{tabular}{@{}p{0.82in}p{1.88in}@{}}
\toprule
Layer / node & Mechanism \;/\; failure it catches \\
\midrule
L1 / UNDERSTAND & code reverse-lookup $\to$ support signals \{type, availability, popularity, works, alias\}; rejects unsupported interpretations. \\
L2 / UNDERSTAND & prompt rules: trust looked-up performer over model memory; $p_{\text{hi}}/p_{\text{lo}}$ gate support; reject niche artists whose works exclude the mention. \\
L3 / PLAN & code guards on query structure: negation $\to$ no positive-ID injection; verb-prefix false match $\to$ drop verb-named items; inject only verified numeric IDs. \\
\bottomrule
\end{tabular}
\end{table}

\textbf{Tool schemas.}
Each tool returns a uniform envelope (tool name, success flag, data, error, latency). The three representative tools below illustrate the heterogeneous tool layer---a production search reused as a retriever, an in-house generative retriever, and a personal-history route. Other routes run in parallel and are merged before REFLECT.

\noindent\textbf{Tool 1: \texttt{catalog\_search}} (multi-index catalog search).
\begin{lstlisting}
{
  "parameters": {
    "name|artist|tags|ip": "string or array",
    "exclude": "array (negation filter)",
    "bpm": "range (tempo filter)",
    "limit": "int, 1..50"
  },
  "returns": "[Song] via {tool_name, success, data, error, latency_ms}",
  "on_empty": "low coverage -> REFLECT may issue retry_relaxed"
}
\end{lstlisting}

\noindent\textbf{Tool 2: \texttt{generative\_retriever}} (Semantic-ID NL$\to$tracks, long-tail tolerant).
\begin{lstlisting}
{
  "parameters": {
    "prompt": "natural-language request",
    "user_id": "string",
    "max_songs": "int (target size N)"
  },
  "returns": "[Song] (semantic-ID matched, playable)",
  "note": "tolerant recall path for long-tail requests"
}
\end{lstlisting}

\noindent\textbf{Tool 3: \texttt{liked\_songs}} (personal liked-set route).
\begin{lstlisting}
{
  "parameters": {
    "user_id": "string",
    "time|filter": "optional window or constraint",
    "pool_size": "int"
  },
  "returns": "[Song] from user's liked set",
  "guard": "hard-protected -> proceed (no relaxation)"
}
\end{lstlisting}

\section{Example Retry Recoveries}
\label{app:cases}

Three triggered sessions from the evaluation set (two preserved, one degraded) are shown as structured traces; \textbf{[retry]} marks the REFLECT-to-PLAN back edge.

\begingroup
\scriptsize\renewcommand{\arraystretch}{0.95}\setlength{\tabcolsep}{2pt}
\begin{table}[H]
\caption{Three triggered retry sessions as structured traces (Cases 1--2 \emph{preserved}, Case 3 \emph{degraded}); the user request appears in each case heading.}
\label{tab:cases}
\centering
\begin{tabular}{@{}p{0.16in}p{0.42in}p{1.28in}p{0.92in}@{}}
\toprule
\# & Node & Thought / utterance & Tool call \& result \\
\midrule
\multicolumn{4}{@{}l}{\emph{Case~1 --- ``nap playlist---no instrumentals (lyrics required),}}\\
\multicolumn{4}{@{}l}{\emph{\quad soothing, no upbeat'' (preserved).}}\\
1 & PLAN & parse 3 constraints; tags=\{soothing\}, exclude=\{instrumental, upbeat\} & \texttt{catalog\_search} $\to$ few candidates \\
2 & REFLECT & coverage gate fails; cause = over-tight tags & \textbf{[retry\_relaxed]} \\
3 & PLAN & broaden soothing tags, keep both exclusions & \texttt{catalog\_search} $\to$ 20 tracks \\
4 & SYNTH & assemble (e.g.\ Khalil Fong, JJ Lin, Zhao Lei) & no instrumental/upbeat leakage \checkmark \\
\midrule
\multicolumn{4}{@{}l}{\emph{Case~2 --- ``1980s Japanese City Pop, female, slow, BPM$<$80'' (preserved).}}\\
1 & PLAN & 4 stacked constraints (era, genre, gender, bpm) & \texttt{catalog\_search} $\to$ 8 candidates ($<\tau$) \\
2 & REFLECT & coverage gate fails; all 4 constraints jointly over-tight & \textbf{[retry\_relaxed]} \\
3 & PLAN & relax BPM ceiling, broaden city-pop tags; keep era/female & add gen. retriever $\to$ 20 tracks \\
4 & SYNTH & Mariya Takeuchi, Anri, Akina Nakamori, Minako Yoshida & era/genre/gender preserved \checkmark \\
\midrule
\multicolumn{4}{@{}l}{\emph{Case~3 --- ``2025 K-pop for group dance, EXO-like but not EXO'' (degraded).}}\\
1 & PLAN & 6 constraints incl.\ year=2025, style=EXO-like, exclude=EXO & \texttt{catalog\_search} $\to$ few candidates \\
2 & REFLECT & coverage fails; year + style constraints too narrow & \textbf{[retry\_relaxed]} \\
3 & PLAN & drop year, relax ``EXO-like'' to ``danceable K-pop'' & \texttt{catalog\_search} $\to$ danceable K-pop \\
4 & SYNTH & BLACKPINK, ITZY, IVE (playable) & release-year + style dropped $\times$ \\
\bottomrule
\end{tabular}
\end{table}
\endgroup

\end{document}